\documentclass[pra,twocolumn,nofootinbib,floatfix]{revtex4}
\usepackage{graphics}
\usepackage{epsfig}
\usepackage{amsfonts}
\usepackage{bm}

\begin{document}
\title{Infrared electron modes in light deformed clusters}
\author{V.O. Nesterenko$^{1,2,5}$, P.-G. Reinhard$^{3}$,
W. Kleinig$^{1,4}$, and D.S. Dolci$^{1}$}
\affiliation{
   $^{1}$
   BLTP, Joint Institute for Nuclear Research, Dubna, Moscow region,
   141980, Russia}
\affiliation{
   $^{2}$Max Planck Institute for Physics of Complex Systems,
   D-01187, Dresden, Germany}
\affiliation{
   $^{3}$Institut f\"ur Theoretische Physik,
   Universitat Erlangen, D-91058, Erlangen, Germany}
\affiliation{
   $^{4}$Technische Univirsitat Dresden, Inst. f\"ur Analysis,
   D-01062, Dresden, Germany}
\affiliation{
   $^5$ E-mail: nester@thsun1.jinr.ru}

\begin{abstract}
Infrared quadrupole modes (IRQM) of the valence electrons in light deformed
sodium clusters are studied by means of the time-dependent local-density
approximation (TDLDA). IRQM are classified by angular momentum components
$\lambda\mu =$20, 21 and 22 whose $\mu$
branches are separated by cluster deformation. In light clusters with a
low spectral density, IRQM are unambiguously related
to specific electron-hole excitations, thus giving access
to the single-electron spectrum near the Fermi surface (HOMO-LUMO region).
Most of IRQM are determined by cluster deformation and so can serve as a
sensitive probe of the deformation effects in the mean field. The IRQM
branch $\lambda\mu =$21 is coupled with the magnetic scissors mode,
which gives a chance to detect the latter.  We discuss two-photon processes,
Raman scattering (RS),
stimulated emission pumping (SEP), and stimulated adiabatic
Raman passage (STIRAP), as the relevant tools to observe
IRQM. A new method to detect the IRQM population in clusters
is proposed.
\end{abstract}

\maketitle

\section{Introduction}

The most prominent electron mode in metal clusters is the Mie surface
plasmon \cite{Mie}. Being the doorway to many structural and dynamical
properties, it has been much investigated in the past, for extensive
summaries see e.g. \cite{KreVol,deH93,Bra93,Hab}.  Besides the
dominant Mie plasmon, many other electron modes, both
electric and magnetic,
have been predicted but not yet observed in clusters, for
a brief overview see \cite{Ne_SN}. Similar modes exist in other
many-fermion systems. In particular, almost all of them were identified
experimentally in atomic nuclei, for a review see \cite{Ber}.

The analogy with deformed nuclei suggests that deformed clusters
should exhibit a family of low-energy (infrared) electron modes which
are absent in spherical systems.  Deformed clusters have a partly
filled valence electron shell and so sustain low-energy
one-electron-one-hole ($1eh$) excitations inside this shell.
Being mainly determined by the deformation splitting,
the excitation energies are rather small and typically lie
in the infrared. Since major shells in light clusters cover electron
levels with the same space parity, the infrared modes should have
positive parity. Their softness and deformation dependence suggest
that these should be infrared quadrupole modes (IRQM).
In axially deformed clusters, IRQM are represented by separate
$\lambda\mu =$20, 21 and 22 branches (where $\lambda=2$ stands for
total angular momentum and $\mu$ its azimuthal component).

Our present knowledge on infrared electronic excitations in clusters
is very poor. Even basic IRQM properties (spectrum, collectivity,
evolution with cluster size and deformation, responses to external
fields) are unknown. It is the aim of this paper to investigate
main features of IRQM from a theoretical perspective, hopefully
delivering useful hints for an experimental search.

IRQM in light and in heavy clusters have a different origin
\cite{na118}. In the present paper, we will consider light
clusters or, specifically, free deformed singly-charged
light clusters. Such clusters have essential advantages with respect
to IRQM. First, they have an extremely low spectral density at low
excitation energies, a feature which facilitates a discrimination of
IRQM in two-photon processes. The low spectral density also
reduces the unwanted level broadening (Landau fragmentation,
electron-electron collisions). Beams of size-selected
singly-charged light clusters are readily available. So, these
clusters are well suited for measurements. Second, as is shown below,
IRQM in light clusters are dominated by one $1eh$ component and can
be directly and unambiguously interpreted in terms of the electron
spectrum in the HOMO-LUMO region. So, IRQM can deliver an important
information on the underlying mean field level structure. Since most
of IRQM are induced by cluster deformation, they allow to study the
evolution of the electronic levels with deformation.  And third, there
is a close connection between $\lambda\mu =$21 branch of the IRQM and
the electronic scissors mode (SM) \cite{LS_M1,prl_M1,
pra_M1,epjd_M1}. This gives a chance to observe the SM, at least
indirectly, through IRQM. SM is a universal dipole magnetic orbital mode
peculiar to deformed systems. It was already observed in a variety of
quantum systems but not yet in clusters.

Our calculations were performed within a linearized time-dependent
local density approximation (TDLDA). The ionic background is
approximated by a soft deformed jellium density, for reviews see
\cite{Cal00,Reibook}. The method goes back to early studies on cluster
spectra \cite{Eka84}. It has been successfully applied
to study the dipole plasmon in spherical \cite{Ne_EPJD_98}
and deformed \cite{Ne_AP,Ne_EPJD_02} clusters.
We will consider here Na clusters as the simplest test
case. However, IRQM properties worked out below are of a general
character and so should appear for all other metal clusters with
pronounced shell structure for valence electrons.

IRQM are not accessible by one-photon transitions which excite
exclusively dipole modes.  Thus one has to use
two-photon processes (TPP) where the target state
is populated via an intermediate state by two (absorption and emission)
dipole transitions. The dipole plasmon can serve
as an intermediate state.

Experimental data on IRQM are very scarce and limited to heavy
clusters (e.g. RS measurements of IRQM in embedded silver clusters
were reported \cite{Duval}).  Moreover, typical two-photon techniques
of atomic and molecular spectroscopy are rarely used for
clusters. This can be partly explained by the fact that clusters have
some specific properties (see discussion in Sec.\ref{sec:exp}), which
makes a straightforward application of these techniques
questionable. In particular, the common methods of detection of the
level population, based on the radiative decay, seem not to be
appropriate for clusters. In this paper, we will examine
applicability to clusters of some widespread TPP:
Raman scattering (RS), stimulated emission pumping (SEP), or
stimulated Raman adiabatic passage (STIRAP).
An alternative method for detection of the level
population in clusters will be proposed.

Altogether, the present study has two goals: i) to provide a first
guide on quadrupole electronic excitations in light deformed clusters;
ii) to explore the access to non-dipole electron modes in clusters by
modern two-photon techniques.

The paper is outlined as follows.  Section \ref{sec:model} describes
the theoretical framework and choice of the most suitable clusters.
In section \ref{sec:origin}, a general hierarchy of quadrupole
electron excitations in clusters is presented and the origin of IRQM
is clarified. In section \ref{sec:res}, the IRQM responses in one- and
two-photon processes are compared and the $1eh$ nature of IRQM is
demonstrated.  In section \ref{sec:exp}, the possibility to observe IRQM
in two-photon process (RS, SEP and STIRAP) is discussed.
Conclusions are drawn in section \ref{sec:conc}. In
Appendix A, the dipole matrix elements responsible for the coupling
between the states involved into two-photon process are analyzed.
In Appendix B, the scissor mode and its connection with the
$\lambda\mu=21$ branch of the IRQM are briefly discussed.

\section{Calculation scheme and cluster choice}
\label{sec:model}

\subsection{Basic spectral properties}

The electron cloud is described by the density-functional theory at
the level of the local-density approximation (LDA) for the ground
state and time-dependent LDA (TDLDA) for the excitations, using
actually the functional of \cite{GL}.  The ionic background of the
cluster is approximated by the soft jellium model allowing for
quadrupole and hexadecapole deformation \cite{Mon95b}. IRQM stay in
the regime of small amplitudes. We thus employ linearized TDLDA, often
called the random-phase-approximation (RPA). The actual implementation
in axial symmetry is explained in \cite{Ne_AP}. The reliability of the
method has been checked in diverse studies of the Mie plasmon in
spherical \cite{Ne_EPJD_98} and deformed \cite{Ne_AP,Ne_EPJD_02}
clusters.

The choice of the clusters was dictated by the following reasons:
 i)  The clusters should be small
enough to possess a dilute and non-collective IRQM spectrum.
Only then the spectrum can be resolved and unambiguously
related to the single-particle levels.
 ii)  Since IRQM are mainly induced by cluster deformation,
the clusters with a strong deformation (both prolate and oblate)
are desirable. The simplest case of axial shape is most
suitable for the analysis.
iii) Shape isomers exhibit different single-electron spectra
\cite{Ne_AP,Ne_EPJD_02}, which can result in smearing out the
low-energy spectral lines. The heavier are clusters, the more
isomers they have \cite{Ne_AP,Ne_EPJD_02}. So, the light
clusters with a strictly dominant equilibrium shape are preferable.
Between them, we should choose the clusters whose ground state
and isomers have the similar (prolate or oblate) shape. Thus we will
minimize the blurring IRQM spectra.
 iv) The jellium approximation is not correct for very
small clusters. This establishes a lower limit for
the cluster size.


The cluster choice can be preliminary done by
reviewing the properties of the dipole plasmon
which will serve as intermediate state in a
two-photon process.  Besides, description of the dipole plasmon
allows to estimate accuracy of the model.
Fig. 1 shows RPA results for the dipole plasmon
in light axially deformed Na clusters: prolate Na$^+_{11}$,
Na$^+_{15}$, Na$^+_{27}$ and oblate Na$^+_{7}$, Na$^+_{19}$.  The
shape of each cluster is characterized by equilibrium quadrupole and
hexadecapole deformations, $\delta_2$ and $\delta_4$, obtained by
minimization of the total cluster energy
\cite{Mon95b,Ne_AP,Ne_EPJD_02}. To simulate temperature and other
sources of line broadening, the photoabsorption cross section is
smoothed by a Lorentzian \cite{Ne_AP}.

As is seen from Fig. 1,
the emerging dipole spectra consist basically of two peaks, the weaker one
($\lambda\mu =$10) corresponds to oscillations along the symmetry axis
and the stronger one to the orthogonal mode $\lambda\mu =$11 (in fact,
two identical modes $\mu=\pm 1$). In most of the clusters, these peaks
are well separated by the deformation splitting. The most heavy
sample Na$_{27}^+$ shows some Landau fragmentation of the resonance
peaks (distribution of the collective strength between several RPA
levels indicated by vertical bars). This is the beginning of a trend
which persists towards heavier clusters. The Landau fragmentation
overrules the deformation splitting already at $N\approx 50$
\cite{Ne_EPJD_02}. Thus one should stay in the region $N\leq 20$ where
the spectra are still dilute and the Landau fragmentation is weak.
Within that region, one should pick samples with a large deformation.
The clusters Na$_{7}^+$, Na$_{11}^+$ and Na$_{15}^+$ seem to be the
best candidates. Besides, their isomeric and ground states have the
similar shapes \cite{Kuemmel}, oblate in Na$_{7}^+$ and
prolate in Na$_{11}^+$ and Na$_{15}^+$.
The analysis of IRQM in the next section confirms this cluster choice.

Fig. 1 shows that in general the folded spectra (solid lines) agree
nicely with the experiment (triangles) which signifies the reliability
of the method.  However, the jellium plus RPA description, working
well for the heavier clusters, worsens for lighter ones. In Na$^+_{7}$
and Na$^+_{11}$, the calculated spectra are redshifted and do not
reproduce the structure details.  The discrepancy is due to the fact
that light clusters tend to be more compressed due to larger surface
tension \cite{Mon94b}. But we use here a constant Wigner-Seitz radius
for reasons of simplicity. Besides, the jellium
approximation is certainly too rough for the description of details
in so light clusters.
The results for the smallest samples may not reach
a quantitative level, but they are still useful for qualitative
consideration as will be done here.
Fig. 1 shows that even in the lightest clusters the jellium TDLDA
sufficiently well reproduces the average energy, principle gross-structure,
and magnitude of the deformation splitting of the resonance. Such
accuracy suffices for our present survey.

\subsection{Two photon process}

We consider a two-photon process running via
$\lambda\mu =10$ or 11 branches of the dipole plasmon as
the intermediate states, see Fig. 2.
The branches are assumed to be well separated
by deformation splitting, which is indeed the case
for strongly deformed light clusters (see Fig. 1).
Then the two reaction paths can be disentangled by tuning
the photon frequency. This allows to specify and monitor the
process. For example, if the reaction
runs only via $\lambda\mu =$10 plasmon branch, then
the population of the low-energy $\lambda\mu =$22 mode is
forbidden and only 20 and 21 modes remain to be considered.
If one of the paths is suppressed,
(see discussion for Na$^+_7$ in the next section),
then the reaction can be tuned to another path.
The comparison of the path rates delivers important
information about the structure of the dipole
plasmon and IRQM.



The population of the IRQM is approximately calculated as a
coherent sum of independent two-step processes, each one being a
product of dipole photoabsorption and photoemission:
\begin{eqnarray}\label{eq:TPP}
  \sigma_{\uparrow\downarrow}(2\mu_2 i_2)
  &=&
  \sum_{i_1}
  \sigma^{ab}_{E1\uparrow} (0 \rightarrow 1\mu_1 i_1)
  \sigma^{em}_{E1\downarrow} (1\mu_1 i_1 \rightarrow 2\mu_2 i_2)
\nonumber\\
  &\propto&
  \sum_{i_1}
  \omega_{1\mu_1 i_1} |\langle1\mu_1 i_1|e{\bf r}|0\rangle|^2
\\
 &&\quad
  \cdot \, (\omega_{1\mu_1 i_1}\!-\!\omega_{2\mu_2 i_2})^3
  |\langle2\mu_2 i_2|e{\bf r}|1\mu_1 i_1\rangle|^2 \,.
\nonumber
\end{eqnarray}
Here $|1\mu_1 i_1>$ and $|2\mu_2 i_2>$ are RPA states of the dipole
and quadrupole spectra, respectively. The index $i_1$ runs
over all RPA states with dipole content in the accessible energy
interval.  The photoabsorption (photoemission) dipole matrix elements
in (\ref{eq:TPP}) define the coupling between the ground and dipole
(dipole and quadrupole) states in two-photon processes.
This coupling determines the Rabi frequency, the decisive
value in TPP.

Eq. (\ref{eq:TPP}) follows from the general expression for RS rate
given elsewhere (see, e.g. \cite{Weissbluth})
if one keeps only the Stokes term and neglects the
interference between the neighboring intermediate states. The Stokes
term alone suffices to illustrate the population of IRQM in two-photon
processes. Moreover, already the dipole matrix elements in (\ref{eq:TPP})
provide a solid ground for the analysis.  As for the
interference, it should be weak in light clusters. Indeed, in Fig. 1
the dipole plasmon in clusters with $N\leq 20$ is represented by a few
states separated by large energy intervals. In any case, the
interference mainly leads to smoothing of the response, which can be
taken into account by a reasonable averaging the results.  In the
present paper, we simulate the interference, temperature and other
smearing factors by folding the results by a Lorentzian profile with
an averaging width $\Delta$.  The spectral width is known to increase
with the excitation energy. So, we use $\Delta=$0.25 eV for the
high-energy dipole and quadrupole plasmons (like in our previous RPA
calculations for the photoabsorption spectra
\cite{Ne_EPJD_98,Ne_AP,Ne_EPJD_02}) and $\Delta=$0.1 eV for the
low-energy IRQM.

It worth noting that the photoabsorption and photoemission dipole
matrix elements in (1) have essentially different structure (see
Appendix A
for more detail).  The former is determined by the $1eh$ part of the
dipole operator $e{\bf r}$ and is generally large.  The latter is
given by $1ee$ (electron-electron) and $1hh$ (hole-hole) parts of the
dipole operator and its value can broadly vary, depending on the
structure of the dipole and quadrupole states.

\section{IRQM in the hierarchy of quadrupole excitations}
\label{sec:origin}

The spectrum of the three-dimensional harmonic oscillator provides a
useful sorting scheme for the valence electron levels in metal
clusters \cite{Cle,deH93,Bra93}. The levels are sorted in perfectly
degenerate bunches, the major quantum shells characterized
by the principle quantum number ${\cal N}=0,1,2,...$.
The shells are separated by appreciable energy gaps and every shell
involves only states of the same space parity $\pi=(-1)^{\cal
N}$. This oscillator picture is well fulfilled in light clusters and
provides still a good approximation in medium and heavy ones. In
axially deformed clusters, the levels are characterized by
Nilsson--Clemenger quantum numbers \cite{Cle} $\nu=[{\cal N}n_z
\Lambda ]$ where ${\cal N}$ is the major shell as before, $n_z$ is the
number of nodes along the symmetry axis $z$, and $\Lambda$ is the
projection of the orbital moment onto the same axis.
Following this scheme, excitations of valence electrons are
characterized by the $\Delta {\cal N}$ value, the difference in shell
for the dominant $1eh$ jumps. Quadrupole excitations have even parity
and are collected into the branches $\Delta{\cal N} =0$ and 2. The
excitations with larger $\Delta{\cal N}$ are weak and can be neglected.

  Figure 3 illustrates the hierarchy of quadrupole excitations. The
lower panels show the unperturbed photoabsorption calculated for the
pure $1eh$ states ignoring the residual interaction. The upper panels
give the RPA photoabsorption.
Let us first look at the strong quadrupole resonance appearing at high
frequencies in the range 2-4 eV.  It is blue-shifted by the residual
interaction and the larger the cluster, the stronger the shift
\cite{Rei96b}. In heavy clusters this resonance is associated with
the quadrupole plasmon. The
resonance is mainly formed by E2 transitions over two major shells
($\Delta {\cal N} =2$).  It exists in clusters of any shape and
exhausts most of the quadrupole strength.  The resonance is energetically
very close to the dipole Mie plasmon (in a simple estimate $\omega_\lambda
=\omega_{\rm pl}\sqrt{\lambda/(2\lambda\!+\!1)}$ where $\omega_{\rm
pl}$ is the frequency of the volume plasmon \cite{Bra93}). In spite of
the energy overlap with the dominant E1 plasmon, the E2 plasmon may,
in principle, be discriminated by means of angular-resolved electron
energy-loss spectroscopy (AR-EELS) at electron scattering angles
$\sim 6^{\circ}$ \cite{Ger}.

In the present study, we are interested not in the high-energy E2
resonance but in the IRQM which, being dilute and  non-collective,
can deliver important information about the electron single-particle
spectrum near the Fermi (=HOMO) level. IRQM are associated with the
low-energy $\Delta {\cal N} =0$ branch created by E2 transitions
inside the valence shell. Being of the $\Delta {\cal N} =0$ origin,
IRQM can exist only in clusters with partly occupied valence shell,
i.e. in deformed clusters.
In Fig. 3, IRQM reside at 0.5-1.5 eV.  As compared with the E2
resonance, the IRQM spectrum is very dilute. It is represented only by
a few well separated levels. This prevents mixing of $1eh$
configurations by the residual interaction and creation of collective
states.  The IRQM persist to keep their $1eh$ nature and so can be
easily identified as particular $1eh$ configurations.  As is seen from
Fig. 3, the IRQM have very weak quadrupole strength in the photoabsorption.
But they may be accessible in two-photon reactions.

%
Fig. 4
shows single-particle levels and $1eh$
quadrupole transitions inside the valence shells in
Na$^+_7$, Na$^+_{11}$, and Na$^+_{15}$. Following this
scheme, the IRQM excitations in Fig. 3 can be
unambiguously identified as particular $1eh$ states.
They include $\{ [101]-[110]\}_{21}$ in Na$^+_7$,
$\{ [220]-[200]\}_{20}$, $\{ [220]-[211]\}_{21}$,
$\{ [220]-[202]\}_{22}$ in Na$^+_{11}$, and
$\{ [220]-[200]\}_{20}$, $\{ [211]-[202]\}_{21}$,
$\{ [211]-[200]\}_{21}$, $\{ [220]-[202]\}_{22}$
in Na$^+_{15}$.

Electric and magnetic modes with the same $\Lambda^{\pi}$ are known to
mix in deformed systems (see, e.g., \cite{Iud}). The mixture of
electric E21 and magnetic orbital M11 excitations is especially
interesting since it provides access to the orbital M1 scissors mode
(SM) \cite{LS_M1,prl_M1, pra_M1,epjd_M1}.
The properties of this mode are sketched in Appendix B.

\section{Results and discussion}
\label{sec:res}


A variety of IRQM spectral distributions is shown in Fig. 5.
The first line of the figure contains the quadrupole
photoabsorption  for IRQM states with $\lambda\mu =$20, 21 and
22.  Though IRQM cannot be observed in photoabsorption, the latter is
useful in any analysis of electron modes and so is worth to be
considered.  The second line exhibits magnetic dipole strengths for
the magnetic dipole scissors mode (SM). The next two lines show the
two-photon populations (\ref{eq:TPP}) when the TPP runs separately
through $\lambda\mu =$10 and 11 branches of the dipole
plasmon.  The intermediate states in the two-photon process involve
all the RPA dipole states from the plasmon region.

The quadrupole spectra in Fig. 5
can be easily identified by $1eh$ transitions exhibited in Fig. 4.
Most of the transitions connect
the levels arising due to deformation splitting. The corresponding
IRQM are determined by the deformation and vanish at the spherical
shape.

The first panel of Fig. 5
shows that IRQM
are selectively active in the quadrupole photoabsorption. Some modes
($\lambda\mu =$22 in Na$^+_{11}$ and $\lambda\mu =$20 and 22 in
Na$^+_{15}$) have small E2-transition matrix elements and so
are suppressed in this one-photon process. However, as is shown below,
these modes can be detected in TPP.

The calculations show that all the IRQM are almost pure
$1eh$ states, i.e. are dominated by one $1eh$ component
(with the related transition in Fig. 4.
The contributions of the dominant components
typically attain $99-100\%$. Even in $\lambda\mu =$21 states in
Na$^+_{15}$, which most deviate from the $1eh$ nature, the
contribution of the leading $1eh$ component are 92-93$\%$.
This feature is additionally illustrated in Fig. 6
where the unperturbed (the dominant $1eh$ configuration alone)
and RPA IRQM strengths are compared. One sees that in both cases
IRQM spectra and strengths are very similar. Only the $\lambda\mu =$21
states in Na$^+_{15}$ show considerable redistribution of the
strength (keeping, nevertheless, predominantly $1eh$ structure).
This case anticipates the involved picture for heavier clusters
where the collective redistribution of strength is much stronger.


The second panel in Fig. 5
exhibits the photoabsorption cross section for the dipole magnetic SM.
The one-to-one
correspondence between E21 and M11 peaks demonstrates the intimate
connection between these two modes. As is discussed in Appendix B, the
scissors mode is driven by cluster deformation and vanishes in the
spherical case. Two $\lambda\mu =$21 states in Na$_{15}^+$
represent an instructive example. The lower $\lambda\mu =$21 state
determined by $[211] \to [202]$ transition between the members of the
deformation multiplet exhibits an appreciable magnetic dipole strength
and thus should carry a large SM fraction. Instead, the higher
$\lambda\mu =$21 state is determined by a $[211] \to [200]$ transition
which takes place even in the spherical case.  The
deformation is not crucial here. This state favors the E21 field and
so can be treated as an ordinary quadrupole mode. Altogether, it is
seen that deformation-induced IRQM provide access to SM.

The third and fourth panels in Fig. 5
show the TPP
cross-section for the reaction paths via $\lambda\mu =10$ and 11
branches of the dipole plasmon. It is seen that TPP response
considerably deviates from the photoabsorption. Besides, the
population of IRQM crucially depends on the reaction path. For example,
it is negligible for $\lambda\mu =$21
states in Na$_{7}^+$ (path 10) and Na$_{15}^+$ (path 11). Suppression
of the TPP transfer in these cases is explained by destructive
effects in the photoemission dipole matrix element (see discussion in
Appendix A). This matrix element can vary to a large extent, depending
on the particular structure of IRQM and dipole intermediate
states. So, for the efficient population of the particular IRQM, one has
to choose the most optimal TPP path. The microscopic calculations can be
used here as a guide.

Because of the $1eh$ nature of IRQM and their strong dependence on
cluster deformation, they can deliver a valuable spectroscopic
information on the electron levels in the HOMO-LUMO region
as well as on the deformation effects in the cluster mean field.
For example, the measurement for $\lambda\mu =$21 state in Na$_{7}^+$
can provide the energy difference between the Fermi [101] and
particle [110] levels (see Fig. 4). If to deduce the Fermi level
energy from the ionization potential data, one finally gets the
energy of [110] level. Simultaneously we will estimate the deformation
splitting of the active subshell. The analysis will be even more
effective, if to combine the TPP data with the photoemission
and inverse photoemission results for the electron spectra.

\section{Experimental access to IRQM}
\label{sec:exp}

\subsection{General view}

Two-photon processes
are widely used in atomic and molecular physics
(for a comprehensive discussion see \cite{Scoles}). Since atomic
clusters are similar to molecules, it would be natural to use
the same reactions in clusters. However, applications of TPP to
clusters are very scarce. This can be partly explained
by our poor knowledge on non-dipole low-energy electron modes
in clusters, partly by peculiarities of clusters.
In this section we will discuss
applicability of traditional TPP methods to clusters
with a particular accent to IRQM.

As compared with atoms and molecules, atomic clusters have at least
three features essential for TPP.  i) The dominant decay channel in
clusters is usually not radiative. Instead, the levels mainly decay
through Landau fragmentation (dissipation of the collective motion
through surrounding $1eh$ excitation), electron-electron
collisions, and electron-ion coupling.  This property spoils the
radiation-based methods (typical for atomic and molecular
measurements) to detect population of levels in clusters. In this
connection, we will propose an alternative detection method based on
the photoabsorption depletion of cluster beam, which is more suitable
for clusters. ii) Cluster excitations have extremely short
lifetimes (10-20 fs for the dipole plasmon in medium clusters) and
usually are rather broad (see e.g. Fig. 1 for the dipole plasmon). The
short lifetimes inhibit the application of adiabatic transfer methods
while large widths impede maintaining the resonance conditions.
iii) Clusters demonstrate a strong shape isomerism.
Cluster beams are usually  a statistical mix of clusters of a different
shape. This blurs the measured low-energy electron spectra.

Light clusters with their dilute spectrum allow to circumvent some of
these troubles. Indeed, Landau fragmentation in such clusters is very
weak, the levels not so broad and the lifetimes are much
longer. Light clusters are strictly dominated by one equilibrium shape
and their isomeric shapes are close to the ground state one \cite{Kuemmel}.
So, we do not expect a strong blurring of the low-energy spectra.
Beams of size-selected singly-charged light clusters are readily
available. The light deformed singly-charged clusters considered in
this paper seem to be most suitable for TPP measurements.
The cluster temperature $\sim$ 100 K could be optimal.
Then the thermal broadening of electron levels is small and,
at the same time, the effects of the electron-ion coupling are yet
smoothed.

In the next subsections we will consider some typical two-photon
processes (RS, SEP, STIRAP)
and estimate their ability to observe IRQM.

\subsection{Raman scattering}

RS is one of the simplest TPP. In this reaction,
a dipole laser-induced transition to an intermediate
electronic level (real or virtual) is followed by
the dipole fluorescence to low-energy levels.

RS measurements of electron infrared modes in
clusters are very rare. We know only one experiment
where  IRQM were observed in heavy silver clusters
embedded into amorphous silica films \cite{Duval}.
The dipole plasmon was used as a resonant intermediate
state.
The further measurements revealed that IRQM are affected by cluster
deformation \cite{Duval_privat}.  These observations show that the
radiative decay of the dipole plasmon is detectable in spite of the
strong competition with other decay channels. This message is very
encouraging for application of TPP to clusters.

RS generally assumes that the coupling between the
intermediate dipole and final quadrupole states is
of the same order of magnitude as the coupling between the
intermediate and ground states. Only then the final state is
successfully populated. This means that the
absorption and emission dipole matrix elements
defined in (\ref{eq:d01}) and (\ref{eq:d12}) should
be of the same scale. The calculations show that this is indeed
the case: in the clusters considered in the present paper
most of the absorption and emission dipole matrix elements
lie inside the interval 2-10 $ea_0$ (in atomic units), i.e.
are basically of the same order of magnitude
(for exception of  some emission matrix elements
strictly suppressed due to destructive contributions,
see discussion in Appendix \ref{sec:A}).
This means that IRQM in light deformed clusters
have a chance to be observed in the resonant
RS (running via the dipole plasmon).

RS is effective only if the coupling between
intermediate and desirable target levels is strong enough.
As was shown for Na$^+_7$, this is not always the case.
To overcome this trouble, we should consider the methods
with stimulated emission into a specific  target level
(SEP, STIRAP).

\subsection{Stimulated emission pumping}

Unlike RS, this method  exploits two lasers, pump
and dump \cite{SEP}. The pump laser is responsible for the
first photoabsorption step. The dump pulse follows the pump
one with some delay and couples intermediate to target states.
If the difference between the pump and dump frequencies
is in resonance with the frequency of the target state,
then the dump radiation stimulates the emission to this
state.

SEP enjoys widespread application for atoms and molecules
but can encounter troubles for clusters where SEP
methods to detect the population (based on measurement
of spontaneous photoemission
or photoelectron yield from certain levels excited by
probe lasers) can fail for short-lived cluster levels
with their small radiative width. In this connection,
we propose a new detection method which does not need
any probe radiative procedure.

The idea is to use depletion of cluster beam caused by
the photoabsorption. It is known that photoabsorption heats
the clusters leading to subsequent evaporation of atoms.
The recoil effect from the evaporated atom forces
the cluster to leave the beam. Thus the photoabsorption
can be measured in terms of depletion. In the scheme we propose,
the frequency of the pump laser is fixed in resonance with
the dipole plasmon while the frequency of the dump laser scans
to get the
resonance with IRQM at
$\omega_P-\omega_D=\omega_{IRQM}$.
Then one can  detect the IRQM population as a dark resonance
in the beam depletion. Indeed if the dump pulse
is not in the resonance with IRQM, then a usual photoabsorption
induced by the pump laser and additionally supported by the dump
laser takes place. In the case of the resonance with IRQM,
part of the energy will be used not for cluster heating
but for stimulated radiative decay from the dipole plasmon
to the given IRQM. So, the photoabsorption (depletion) should
demonstrate a sharp decrease, kind of a dark resonance. The deeper
the dark resonance, the more populated is the IRQM. The photoionization
channel plays no role because the ionization potential in light clusters
is higher than $\omega_P+ \omega_D$. If the laser intensities are
not too high, then  multi-photon ionization can be also
neglected.  The scheme is simple and suitable for clusters.

However, even in the best case, SEP can transfer to the target
level only 25$\%$ of the population \cite{SEP}. Spontaneous emission
from the intermediate level results in an additional leaking.
So, it is worth to continue our discussion and consider more
efficient methods.
\subsection{STIRAP}

Much better results can be obtained with STIRAP
\cite{Berg_Shore,Berg} which promises up to 100$\%$
of the population transfer from the initial to the target
level. So high effectivity is provided by the coherent
character of the process and the principle possibility to avoid
the leaking from the intermediate levels. Like the previous method,
STIRAP also implements two lasers: pump laser to couple the initial
state  $|0>$  and  intermediate state $|1>$ and Stokes laser
to stimulate the emission from $|1>$  to the target state $|2>$.
However, STIRAP is more involved because it implies
the coherent adiabatic transfer.

STIRAP has three principle requirements:
i) two-resonance condition $\omega_P-\omega_S=\omega_2-\omega_0$
which allows a detuning
$\Delta=(\omega_1-\omega_0)-\omega_P=(\omega_1-\omega_2)-\omega_S$
from the intermediate state frequency;
ii) counterintuitive sequence
of the pulses when the Stokes pulse proceeds the pump one;
3) adiabatic evolution.

Under these conditions, one of the time-dependent
eigenfunctions of the system
is a superposition of initial and target bare states
only: $|b_0(t)>=c_0(t)|0>+c_2(t)|2>$.
It is reduced to $|0>$ at the beginning of
the adiabatic evolution and to $|2>$ at the end.  Hence,
the system finally finds itself in the target state.
The intermediate state $|1>$ is not involved to  $|b_0(t)>$
and so is not populated at any time. Thus any leak in the
population is avoided and the transfer is complete.
The main point is to evolve the system adiabatically,
keeping it all the time  in the state $|b_0(t)>$.
STIRAP is widely used in atomic and molecular
spectroscopy. It is rather
insensitive to precise pulse characteristics.
Both continuous and pulse lasers can be implemented.

Let's examine the STIRAP requirements for atomic clusters,
in particular for the  IRQM population via the dipole plasmon.

{\it Two-resonance condition}.
This condition can be obviously maintained even
for a broad dipole plasmon. However, the role of
detuning $\Delta$ (which is used as an additional tool
to prevent the $|1>$ admixture to $|b_0(t)>$)  becomes
vague. The dipole plasmon vanishes slowly at its
flanks and  so rather large detuning is necessary.
This can be done even under the restriction
$\Delta \ll \omega_P \pm \omega_S$ imposed by the rotation
wave approximation. At the same time, this point
was not yet properly investigated and, probably,
some fraction of $|1>$ will admix $|b_0(t)>$. Then
the population will not be complete, although it still can
remain large.

{\it Counterintuitive order of pump and Stokes pulses}.
The order when Stokes pulse precedes the pump one is crucial
for the maximal population \cite{Berg_Shore,Berg}. Then the
Stokes pulse prepares the coherent superposition of the
intermediate and final states just before arriving the pump
pulse to stimulate the subsequent desirable transfer.
Stokes and pump pulses must overlap and the overlapping time
$\Delta \tau$ determines duration of the adiabatic
evolution. This time should not be longer than the
lifetime of the dipole plasmon, i.e. should not exceed
10-100 fs. Though  such $\Delta \tau$ is extremely short,
it is quite accessible in modern experiments \cite{SEP}.

{\it Adiabatic passage}. This condition is most tough.
Following \cite{Berg_Shore,Berg} it can be formulated as
\begin{equation}\label{eq:adia}
 \Omega \Delta\tau > 10
\end{equation}
where $\Omega=\sqrt{\Omega_P^2+\Omega_S^2}$
is the average of the pump and Stokes Rabi frequencies.
The Rabi frequency is
\cite{Berg_Shore}
\begin{equation}
 \Omega = \frac{|d|}{\hbar}\sqrt{\frac{2I}{c\epsilon_0}}
 \simeq 2.20 \cdot 10^8 \; |d[ea_0]| \; \sqrt{I[\frac{W}{cm^2}]}
 \; s^{-1} \; ,
\end{equation}
where $d$ is the dipole coupling matrix element in atomic units,
$I$ is the laser intensity in $W/cm^2$, $c$ is the light speed,
$\epsilon_0$ is the vacuum dielectric constant.
It is easy to estimate (for $d\sim 5 ea_0$ and $\Delta\tau =$10 fs)
that the condition (\ref{eq:adia}) is maintained
for intensities $I > 10^{12}$ $W/cm^2$.
At so high intensities, the multiphoton ionization
and fragmentation of clusters should take place
\cite{Cal00}, which can spoil STIRAP.

However, there is a way to circumvent the trouble.
The condition (\ref{eq:adia}) was obtained for {\it one} intermediate
level. But the realistic spectrum of the dipole plasmon consists
of a sequence of dipole levels. In this case, the STIRAP
condition should be revised.  The realistic case is more
complicated but, at the same time, opens new possibilities
for STIRAP. In particular, one may loose the adiabatic STIRAP
condition and thus the requirements for the
laser intensity.

A general case of N intermediate states, each with its own
coupling and detuning was studied in \cite{Vitanov}.
It was shown that the trapped adiabatic state $|b_0(t)>$ can
be created only when the ratio between each pump coupling and the
respective Stokes coupling is the same for all intermediate
states. Following our calculations, this condition is
unrealistic for atomic clusters. However, the softer alternative
adiabatic requirements can be  formulated.
In particular, in the general case of arbitrary couplings,
one may tune the pump and Stokes lasers just below all
intermediate states and thus form so called adiabatic-transfer
state which also results in a high if not complete
population of the target level. Unlike $|b_0(t)>$, the
adiabatic-transfer state can have admixtures from the
intermediate states during the evolution period $\Delta \tau$
and so some population leaking is unavoidable.
Nevertheless, we have here a solid adiabatic
transfer with a high population of the target state.
Finally, one may conclude that STIRAP can be used for the
efficient population of IRQM in clusters.

\section{Conclusions}
\label{sec:conc}

We have presented a first exploration of infrared quadrupole electron
modes (IRQM) in light deformed clusters.  Most of IRQM are induced by
cluster deformation and thus can deliver useful information on
deformation effects, e.g. on the deformation splitting of electron
levels. Besides, IRQM are about pure one-electron-one-hole excitations
and so give access to the single-electron spectrum. We
explained the origin of IRQM and showed that they can be
easily identified in the dilute spectrum of light clusters.

In the second part, we examined some typical two-photon
processes (Raman scattering, stimulated emission population
and stimulated adiabatic Raman passage) which are widely used
in atomic and molecular spectroscopy but not yet for atomic
clusters. It was shown that, in spite of
some peculiarities of clusters (broad resonances,
short level lifetimes, domination of non-radiative decay channels),
these TPP can be applied to populate IRQM. Besides,
a new method to detect the population of the target cluster states
was proposed. TPP measurements of IRQM
can be supplemented by photoemission and inverse photoemission
experiments delivering the similar information.
We hope that our analysis will stimulate application of the
experimental TPP methods of atomic and molecular spectroscopy
to atomic clusters.

Non-dipole electron excitations in clusters represent
a new promising research field. They deliver interesting physics and can
serve as a robust test for the theory which still paid main attention
to integral characteristics of clusters but not to so fragile patterns
as the single-electron spectra.
It worth noting that the single-electron spectra are very sensitive
to many cluster features and so can be used as an effective tool
for investigation of these features.

Though our calculations have been done for sodium clusters,
the qualitative results we obtained are of a general nature and
should be valid for other metal clusters as well. In particular,
similar deformation-induced IRQM are expected for
supported clusters where they can serve as sensitive
indicators of the interface interaction and cluster deformation.
Besides, IRQM can be important for the quantum transport in
clusters where they can lead to a resonant transmission.

\begin{acknowledgments}
The work was supported by the Visitors Program of Max
Planck Institute for the
Physics of Complex Systems (Dresden, Germany). We thank
professors E. Duval and J.-M. Rost
for valuable discussions.
\end{acknowledgments}

\appendix
\section{Dipole matrix elements}
\label{sec:A}

The dipole and quadrupole electron states are described as
RPA modes
\begin{equation}
  Q_{\lambda\mu i}^{\dag}=\frac{1}{2}\sum_{eh}
(\psi^{\lambda\mu i}_{eh}a^{\dag}_e a_{h}
-\phi^{\lambda\mu i}_{eh}a_e a^{\dag}_{h})
\label{eq:Q}
\end{equation}
where
\begin{equation}
\psi^{\lambda\mu i}_{eh}
\sim N_{\lambda\mu i}
\frac{f^{\lambda\mu}_{eh}}{\epsilon_{eh}-\omega_{\lambda\mu i}} \: ,
\quad
\phi^{\lambda\mu i}_{eh}
\sim N_{\lambda\mu i}
\frac{f^{\lambda\mu}_{eh}}{\epsilon_{eh}+\omega_{\lambda\mu i}}
\end{equation}
are forward and backward
amplitudes characterizing contributions of electron-hole
($1eh$) configurations to the mode. Furthermore,
$N_{\lambda\mu i}$ is the normalization coefficient,
$f^{\lambda\mu}_{eh}$ is the single-particle matrix element
of the residual two-body interaction, $\epsilon_{eh}$ and
$\omega_{\lambda\mu i}$ are energies of the $1eh$ and RPA
excitations, respectively. RPA describes equally well both
collective  and non-collective modes. In the latter case
(i.e. in the limit $|\lambda\mu i> \to |1eh>$),
only the amplitude $|\psi^{\lambda\mu i}_{eh}| \to 1$
survives while all others vanish.

The photoabsorption matrix element responsible for the
coupling between the ground and dipole states
is determined  by $1eh$ part of the dipole operator
and reads
\begin{equation} \label{eq:d01}
 \langle 1\mu_1 i_1|erY_{1\mu}|0> = \sum_{eh}
f^{E1\mu_1}_{eh}
(\psi^{1\mu_1i_1}_{eh}+\phi^{1\mu_1i_1}_{eh})
\end{equation}
where $f^{E1\mu_1}_{eh}$ is the  matrix element of
the dipole transition. If to put
$f^{1\mu_1}_{eh} \approx f^{E1\mu_1}_{eh}$
and to take into account  that all the relevant dipole
$1eh$-levels are blue-shifted by the dipole interaction, then
\begin{equation}
\langle 1\mu_1 i_1|erY_{1\mu}|0> \approx
2 N_{1\mu_1 i_1}
 \sum_{eh}
\frac{(f^{E1\mu_1}_{eh})^2 \epsilon_{eh}}
{\epsilon_{eh}^2-\omega_{1\mu_1 i_1}^2}
\end{equation}
and it is easy to see that all $1eh$ levels contribute constructively
to the transition.

Instead, the photoemission dipole matrix element
connecting the dipole and quadrupole RPA modes
is determined  by $1ee$ and $1hh$ parts of the dipole
operator and has the more complicated structure:
\begin{eqnarray}\label{eq:d12}
  \langle 2\mu_2 i_2 | erY_{1\mu}| 1\mu_1 i_1 \rangle
\nonumber
\\
 =\frac{1}{2} \lbrack
\sum_{ee'} f^{E1\mu}_{ee'} \sum_{h}
 \langle
\psi^{2\mu_2i_2}_{eh}\psi^{1\mu_1i_1}_{he'}+
\psi^{2\mu_2i_2}_{he'}\psi^{1\mu_1i_1}_{he}
 \rangle
\nonumber
\\
+ \sum_{hh'} f^{E1\mu}_{hh'} \sum_{e}
 \langle
\psi^{2\mu_2i_2}_{he}\psi^{1\mu_1i_1}_{eh'}+
\psi^{2\mu_2i_2}_{h'e}\psi^{1\mu_1i_1}_{eh}
 \rangle \rbrack \; .
\end{eqnarray}
The terms with backward amplitudes,
$\sim \phi\phi $,
are omitted in (\ref{eq:d12}) since they are usually small.

The matrix element (\ref{eq:d12})
becomes simpler when the quadrupole state
is dominated by a single $1eh$ configuration, say
$\{ \bar{e}_2 \bar{h}_2 \}$. As is shown in Sec. \ref{sec:res},
this is indeed a common case.
Then all the quadrupole amplitudes vanish except of
$|\psi^{2\mu_2 i_2}_{\bar{e}\bar{h}}| \to 1$
and Eq. (\ref{eq:d12}) is reduced to
\begin{eqnarray}
  \langle \{ \bar{e}_2 \bar{h}_2 \}|erY_{1\mu}|1\mu_1 i_1\rangle
\nonumber
\\
 =\pm\frac{1}{2}  \lbrack  \sum_e
f^{E1\mu_1}_{\bar{e_2}e} \psi^{1\mu_1i_1}_{\bar{h}_2e}
+ \sum_h
f^{E1\mu}_{h\bar{h}_2} \psi^{1\mu_1i_1}_{\bar{e}_2h}
 \rbrack  \; .
\label{eq:emiss_simpl}
\end{eqnarray}
Now only the dipole amplitudes including the hole
(first term) or particle (second term) from the pair
$\{ \bar{e}_2 \bar{h}_2 \}$ contribute to the transition.
The contribution is considerable only subject to
the large dipole amplitude and strong matrix element.
(The latter take place if the matrix element fulfills
the asymptotic Nilsson selection rules \cite{Nilsson,BM}:
 E1$\mu$ transition between $[{\mathcal N}_i n_{iz} \Lambda_i]$ and
$[{\mathcal N}_j n_{jz} \Lambda_j]$ states is favored if
${\mathcal N}_i={\mathcal N}_j\pm 1, \quad n_{iz}=n_{jz}\pm 1$ for $\mu=0$
and $\quad n_{iz}=n_{jz}$ for $\mu=1$).
Under these tough requirements, only a few terms
yield large contributions to (\ref{eq:emiss_simpl}).
Depending on the structure of the dipole and quadrupole states,
the contributions can have different signs and thus lead to
constructive or destructive results.
So, the magnitude of the photoemission matrix element
(and thus the coupling between the dipole and quadrupole states)
can be large (like for the  photoabsorption) or very small.

\section{Scissors mode}
\label{sec:B}

The SM is a general dynamical phenomenon already found or predicted in
different finite quantum systems (atomic nuclei, metal clusters,
quantum dots, dilute ultra-cold gases of Bose and Fermi atoms),
see the review \cite{epjd_M1}.
All these different systems have two features in common: broken
spherical symmetry (deformation) and a two-component structure.

Like the quadrupole modes, the SM separates
into low-energy ($\Delta {\cal N} =0$) and high-energy
($\Delta {\cal N} =2$) branches. In this paper we consider
only the low-energy branch which is mixed with IRQM.
Macroscopically, this branch is treated as a small-amplitude rotational
oscillations of a spheroid of valence electrons against a
spheroid of the ionic background (hence the name scissors mode).
Like IRQM, the SM is driven by cluster deformation and
its energy scale is naturally determined by the deformation
splitting of the electron levels. In axial clusters
with a quadrupole deformation $\delta_2$,
the energy and magnetic strength of the mode
are estimated as \cite{LS_M1,prl_M1}:
\begin{equation}
\omega = \frac{20.7}{r_{s}^{2}}N_{e}^{-1/3}\delta_{2} \ eV,
\qquad B(M1) \simeq N_{e}^{4/3}\delta_{2} \ \mu _b^{2}
\label{eq:om}
\end{equation}
where
$r_s$ is the Wigner-Seitz radius (in $\AA$), $N_{e}$ is the
number of valence electrons, and $\mu_b$ is the Bohr magneton.
It is seen that both the energy and strength are
proportional to the deformation parameter $\delta_2$ and
vanish for the spherical shape.

In axially symmetric systems, the SM is generated by the orbital
momentum fields, $L_x$ and $L_y$, perpendicular to the symmetry axis
$z$ and, like the quadrupole mode $\lambda\mu=21$,
forms the states $|\Lambda^{\pi}=1^{+}>$.
SM strongly responds to an external magnetic dipole field.
Besides, it determines the van Vleck paramagnetism in deformed
clusters \cite{LS_M1,prl_M1, pra_M1,epjd_M1}. The experimental
search of the SM in clusters is still in very beginning
\cite{Duval}.

There is an intimate connection between
the scissors and quadrupole E21 modes in deformed clusters.
If fact, both SM and E21 mode are parts of a general
motion of multipolarity $\lambda\mu=21$.
To illustrate this point, we expand a single-particle
electron state $\nu$ in terms of a spherical basis  $(n L\Lambda )$
\begin{equation}
 \Psi_{\nu =[{\cal N}n_z\Lambda ]}
 =
 \sum_{nL} a^{\nu}_{nL}R_{nL}(r) Y_{L\Lambda}(\Omega)
\end{equation}
and estimate the SM $1eh$ matrix element:
\begin{eqnarray}\label{eq:me}
 \langle\Psi_{p}|{\hat L}_{x}|\Psi_{h}\rangle
 &\propto&
 \delta^{\mbox{}}_{{\pi}_{p},{\pi}_{h}}
 \delta^{\mbox{}}_{\Lambda_{p},\Lambda_{h}\!\pm\!1}
\\
 &\cdot & \sum_{nL}
 a^{p}_{nL}a^{h}_{nL}\sqrt{L(L\!+\!1)\!-\!\Lambda_h(\Lambda_h\!\pm\!1)}
 \,.
\nonumber
\end{eqnarray}
It is seen that the SM operator connects only components
from one and the same basis level $nL$.
Indeed, the operators ${\hat L}_{x}$ and ${\hat L}_{y}$
have no $r$-dependent part and so, due to orthogonality
of the basis functions $R_{nL}(r)$, cannot connect the
components with different $nL$. But the latter can be done
by the quadrupole operator $r^2 Y_{21}$.
In this sense the SM operator is more selective than E21,
though both operators generate transitions of the same
multipolarity. $\Lambda^{\pi}=1^+$ states
involve both SM and E21 modes and respond to both M11 and E21
external fields. The states are treated as magnetic dipole SM
or electric quadrupole E21, depending on each of the two
responses dominates.

\newpage

{\bf \large FIGURE CAPTIONS}

\vspace{0.5cm}\indent
{\bf Figure 1}:
Photoabsorption dipole cross section
in light axially deformed Na clusters.
Quadrupole and hexadecapole deformation
parameters are indicated in boxes.
RPA results are given as vertical bars (in $eV \AA^2$) and
as a strength function smoothed by the Lorentz weight
with the averaging parameter 0.25 eV.
Separate contributions to the strength function
from the  $\lambda\mu =$10 and 11
dipole branches (the latter is twice stronger) are
shown by dashed curves. The experimental data  (triangles) from
\protect\cite{SH} are given for the comparison.

\vspace{0.5cm}\indent
{\bf Figure 2}:
Two-photon process: scheme of population of IRQM states
$\lambda\mu =$20, 21 and 22 via the $\lambda\mu =10$ (left)
and 11 (right) branches of the dipole plasmon.

\vspace{0.5cm}\indent
{\bf Figure 3}:
Quadrupole strength distribution in light Na
clusters. Lower panels: the unperturbed $1eh$ strength (without
residual interaction). Upper panels:
the RPA strength (with residual interaction).
The results are given as bars (for every discrete RPA or $1eh$ state)
and as smooth strength functions obtained by folding with a Lorentzian
of width $\Delta =$0.25 eV. The IRQM strength (enclosed by
the circles at 0.5-1.5 eV) is very weak and so rescaled by the
factor $10^2$.

\vspace{0.5cm}\indent
{\bf Figure 4}:
The electron level scheme for Na$_7^+$, Na$^+_{11}$ and Na$^+_{15}$
in the spherical limit (left) and at the equilibrium
deformation (right). Occupied and unoccupied states are drawn
by solid and dashed lines, respectively.
The Fermi (HOMO) level is marked by index F.
Arrows depict the possible low-energy
hole-electron $E2\mu$ transitions.

\vspace{0.5cm}\indent
{\bf Figure 5}:
IRQM in light clusters. The plots exhibit quadrupole photoabsorption
(uppermost panels, marked E2), scissors M1 photoabsorption (second
line, marked M1), and two-photon population of IRQM via $\lambda\mu
=10$ (third line, marked TPP E10) and 11 (lowest panels, marked TPP
E11) dipole branches. IRQM are depicted by solid ($\lambda\mu =20$),
dashed ($\lambda\mu =21$), and dotted ($\lambda\mu =22$) curves. The
strengths are smoothed by a Lorentzian with $\Delta =0.1$ eV.

\vspace{0.5cm}\indent
{\bf Figure 6}:
Quadrupole strengths for particular IRQM. The unperturbed
$1eh$ (dashed curve) and RPA (solid curve) strengths
are compared. The results are smoothed by a Lorentzian
with the width $\Delta =0.1$ eV.

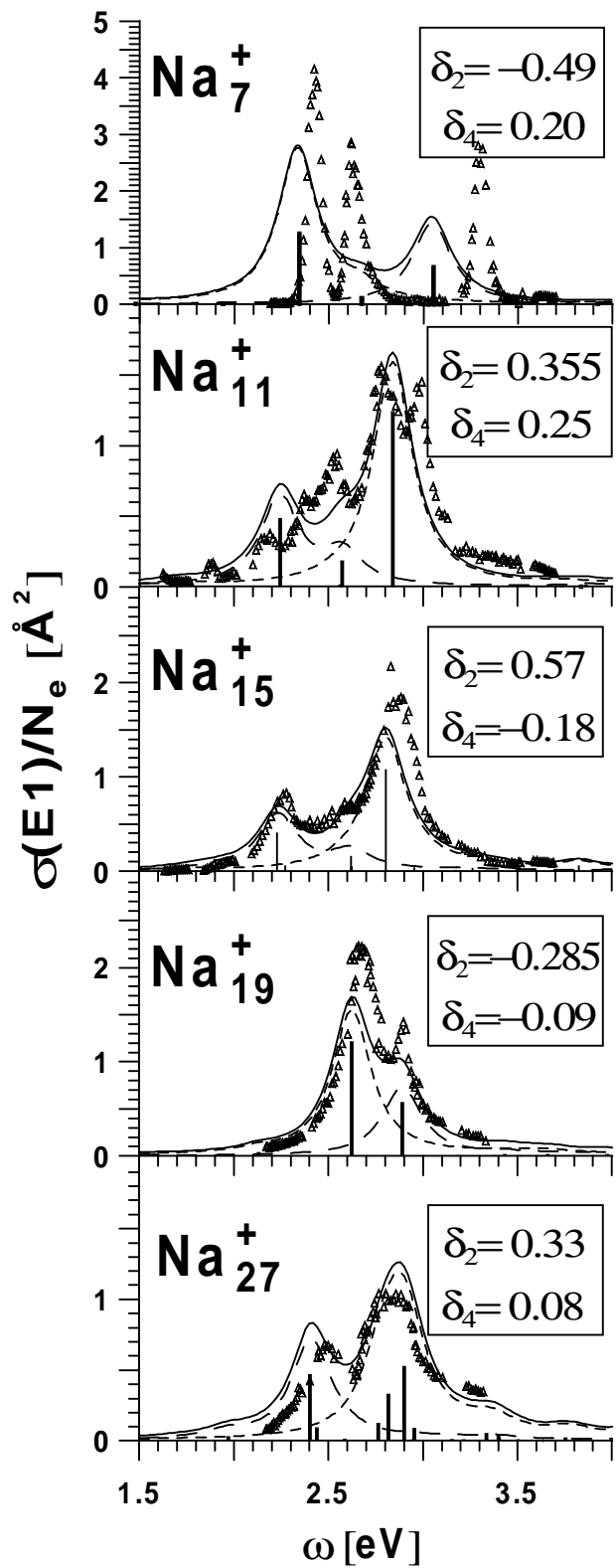
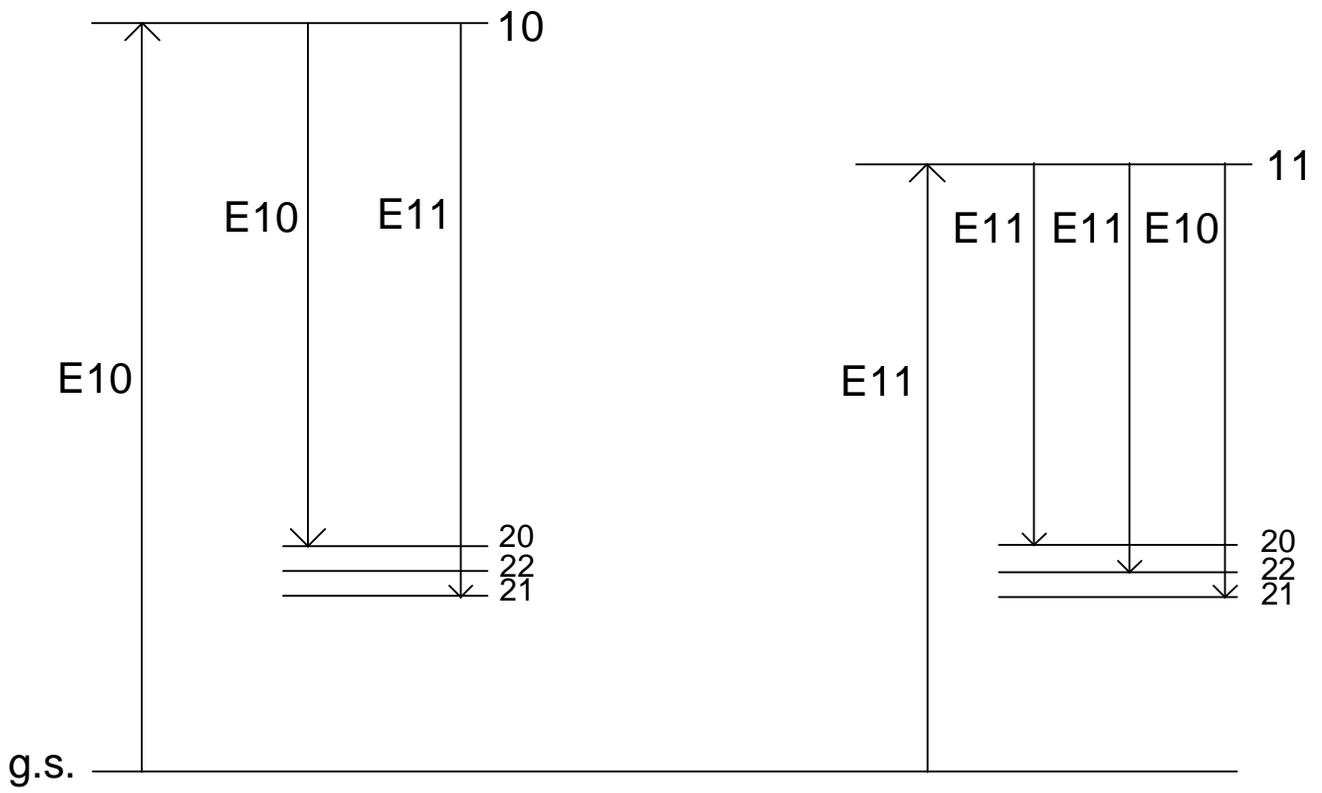
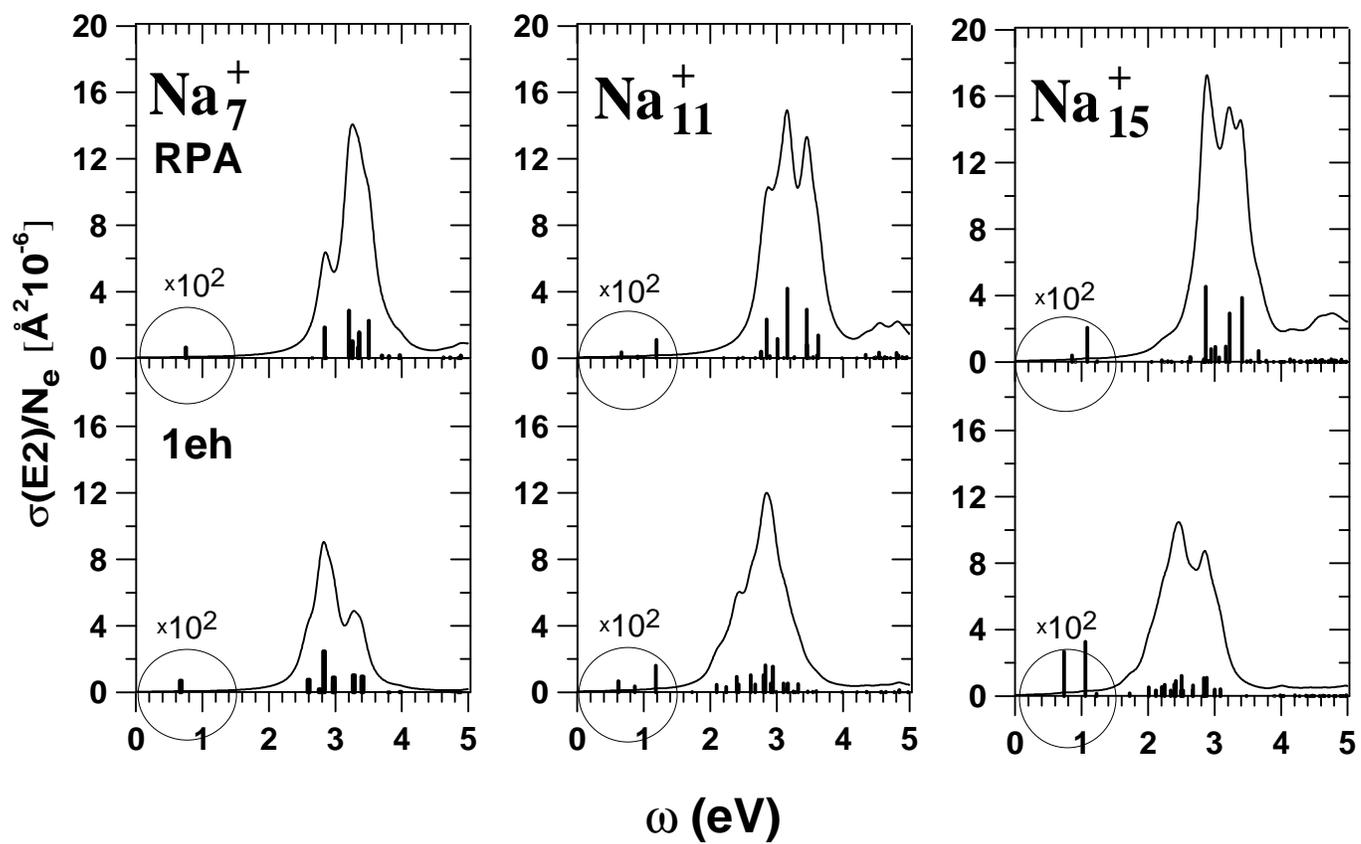
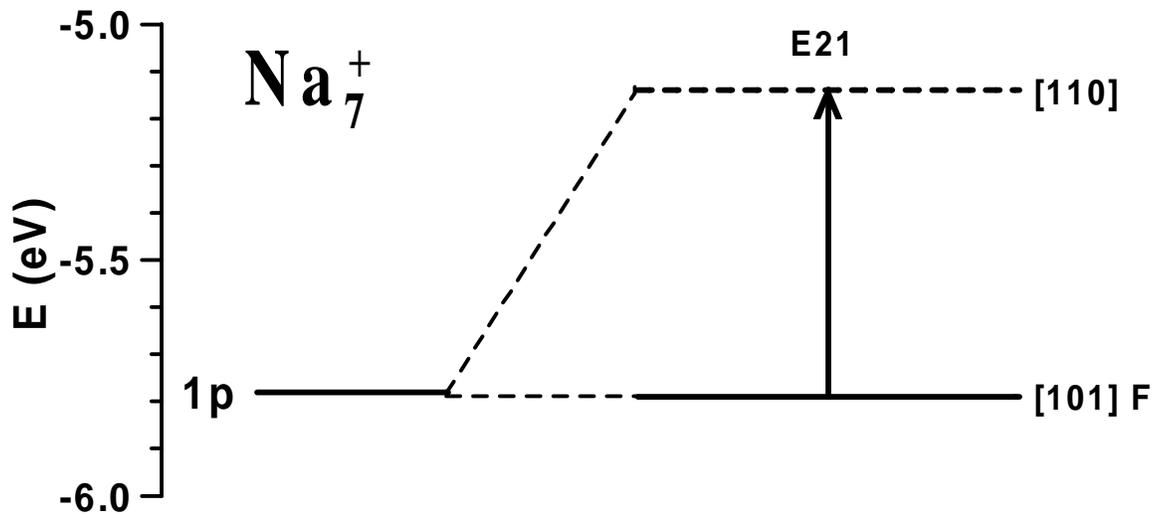
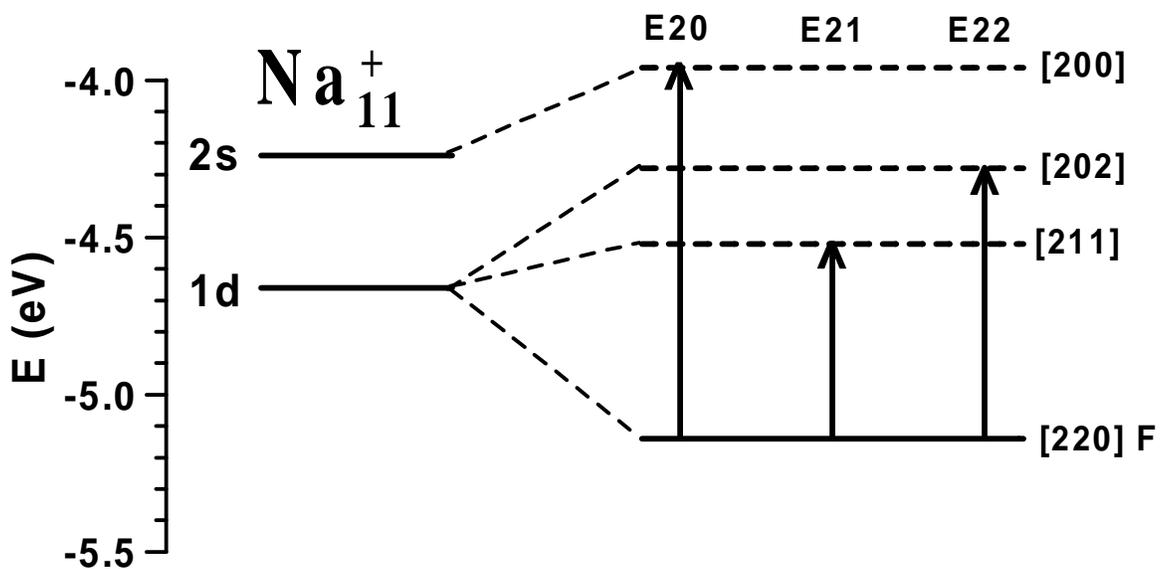
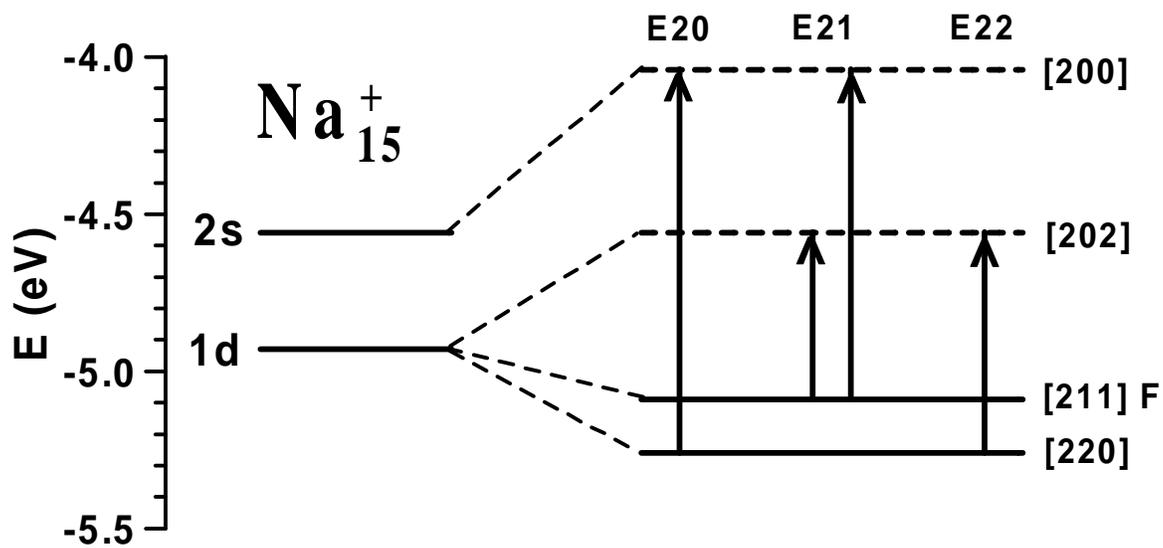
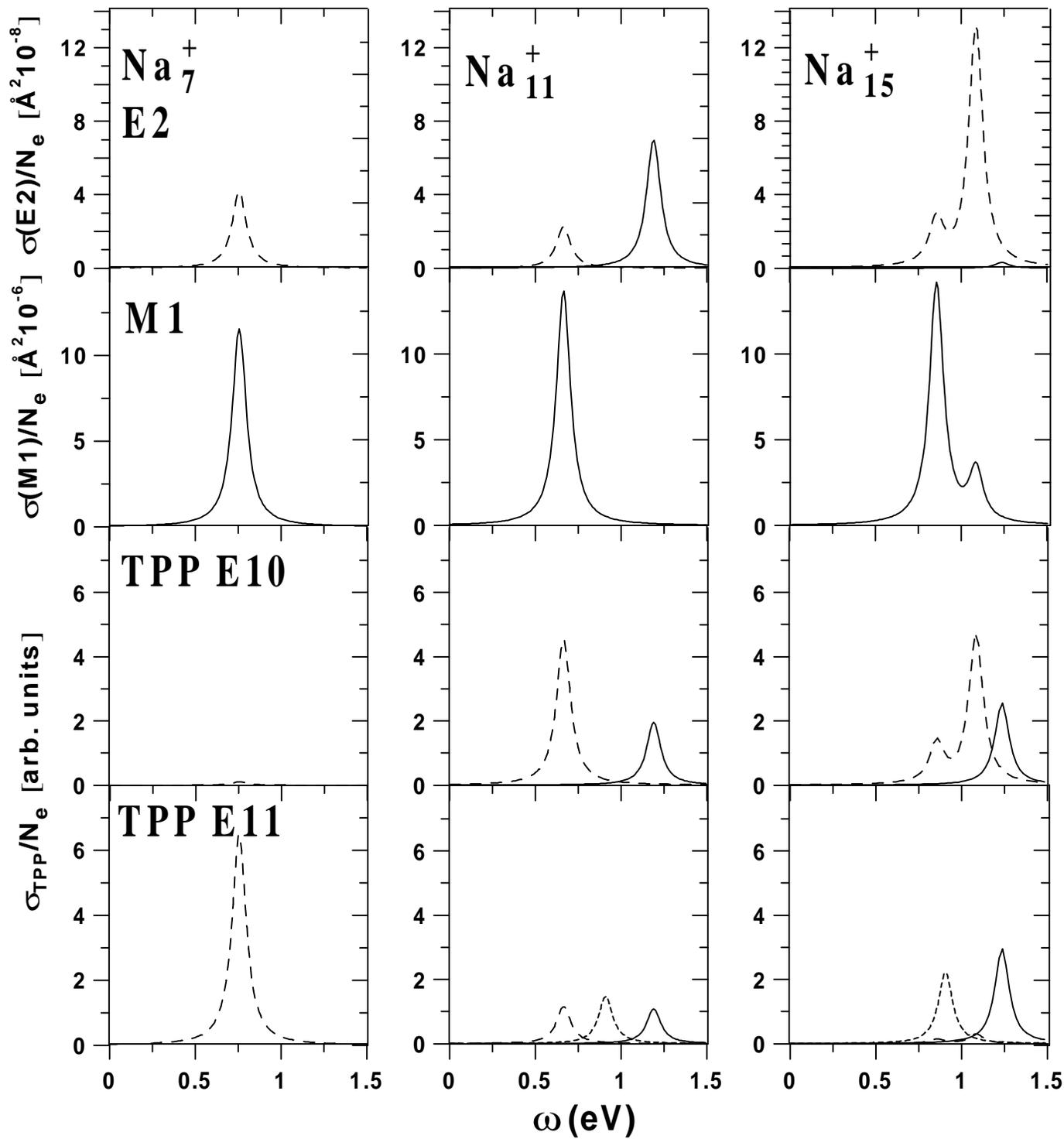
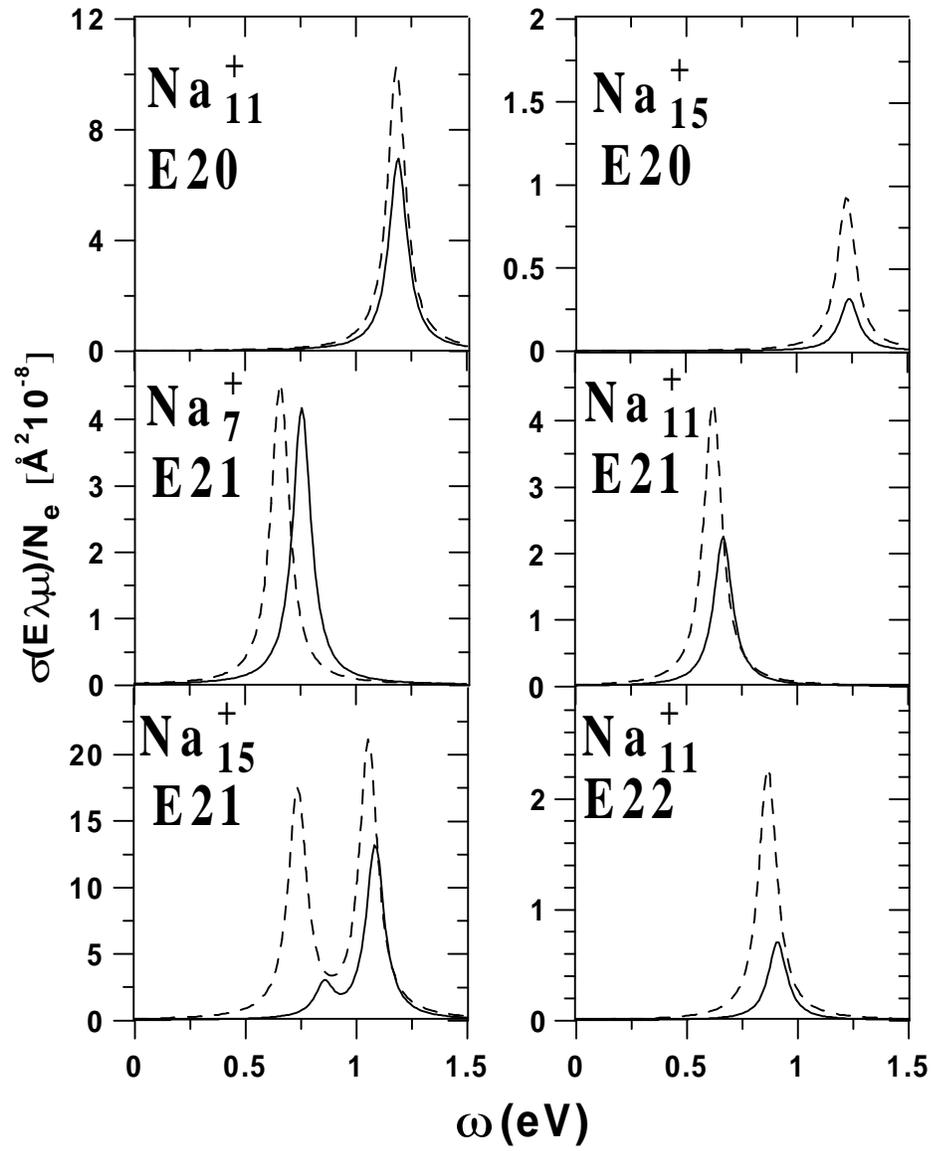
\end{document}